# High-resolution electro-optically sampled broadband dual-comb spectroscopy across mid-IR to terahertz at video rate

Dmitrii Konnov[1], Andrey Muraviev[1], Sergey Vasilyev[2], and Konstantin Vodopyanov[1,a]

1. CREOL, College of Optics and Photonics, University of Central Florida, Florida 32816, USA
2. IPG Photonics Corporation, 377 Simarano Drive, Marlborough, MA 01752, USA

a) *vodopyanov@creol.ucf.edu*

**ABSTRACT**

Ultrabroadband electro-optic sampling with few-cycle optical pulses is known to be an extremely sensitive technique to detect electric field amplitudes. By combining this method with dual-comb spectroscopy and with a new class of ultrafast lasers, we perform high-resolution (< 10 MHz, 0.0003 cm$^{-1}$) spectroscopic measurements across the whole frequency range of 1.5 to 45 THz (6.6–200 µm) with an instantaneous spectral coverage exceeding an octave (e.g., 9–22 µm). As a driving source, we use a pair of highly mutually-coherent low-noise frequency combs centered at $\lambda \approx$ 2.35 µm produced by mode-locked solid-state Cr:ZnS lasers. One of the two combs is frequency downconverted via intrapulse difference frequency generation to produce a molecular 'sensing' comb, while the second comb is frequency doubled to produce a near-IR comb for electro-optic sampling (EOS). An ultra-low intensity and phase noise of our dual-comb system allows capturing a vast amount of longwave spectral information (>200,000 comb-mode spectral lines) at up to video rate of 69 Hz and with the high dynamic range limited by the shot noise of the near-IR EOS balanced detection. Our long-wavelength IR measurements with low-pressure gases: ethanol, isoprene and dimethyl sulfide reveal spectroscopic features that had never been explored before.

## I. INTRODUCTION

The technique of dual-comb spectroscopy (DCS) [1,2] has been rapidly expanding over the last two decades starting from the proof-of-concept work [3] where the key advantages of this method over the traditional Fourier transform infrared spectrometry were revealed, namely broadband coverage combined with high spectral resolution, high acquisition speed, high precision, and the absence of moving parts. Mid-infrared (mid-IR) spectral region (3–25 µm) is of special interest for molecular spectroscopy and trace molecular detection since molecules have their strongest absorption bands across this range. Significant progress in generating extremely broadband mid-IR frequency combs became possible due to the new development of mode-locked fiber [4] and solid-state [5,6,7,8] laser combs and efficient downconverting their frequencies through optical parametric oscillation (OPO) [9,10,11,12], difference-frequency generation (DFG) [13,14], and intra-pulse DFG (IDFG) [15,16] based on advanced $\chi^{(2)}$ nonlinear crystals [17] (see also [18,19] and references therein). Simultaneously, great effort has been made in developing chip-scale frequency combs based on microresonators and waveguides [20,21,22,23], quantum cascade lasers [24,25,26], and interband cascade lasers [27].

More recently, a system has been reported that for the first time has demonstrated all the advantages of the dual-comb method *simultaneously*, namely: broadband instantaneous spectral coverage (6.6-11.4 µm), superior resolution (<0.0027 cm$^{-1}$) and high detection speed (10 Hz) based on efficient downconversion of phase-locked 2.4-µm combs to the longwave IR (LWIR) domain via IDFG in zinc germanium phosphide (ZGP) crystals and acquisition of interferograms using a fast liquid nitrogen cooled HgCdTe photo-detector [28]. However, reaching to >12 µm wavelengths remains a major challenge for photon detectors, as they suffer from higher noise and slower response at long infrared wavelengths, even when operating at cryogenic temperatures.

Kowligy et al. [29] presented a new approach to DCS that combines the IDFG method to create a LWIR 'sensing' comb and electro-optic detection using a near-infrared (NIR) 'sampling' comb and eliminates the need for cryogenic IR detectors. Essentially, electro-optic sampling (EOS) combines three attractive techniques for low-noise detection of LWIR radiation: up-conversion from LWIR to NIR frequencies, optical time gating that eliminates background noise, and heterodyning that potentially allows quantum-limited LWIR detection [30,31,32].

Here we report a novel approach to EOS-DCS using mode-locked Cr:ZnS lasers as the driving source. These lasers have emerged as longwave alternatives to Ti:Sapphire technology, offering several advantages including efficient pumping schemes and the highest LWIR downconversion efficiency. With this approach, we were able to conduct spectroscopic measurements spanning the entire frequency range from 1.5 to 45 THz (corresponding to wavelengths of 6.6-200 μm) with an instantaneous spectral coverage of up to an octave, absolute frequency referencing, and the capability to resolve hundreds of thousands of comb-mode lines at video rate.

## II. EXPERIMENTAL SETUP

### A. Driving laser combs at 2.35 μm

The front end of our DCS system is a pair of Cr:ZnS laser frequency lasers (Fig. 1, inset), each laser consisting of a polycrystalline Cr:ZnS master oscillator pumped by an Er-doped fiber laser (EDFL) at 1567 nm wavelength, and a single-pass Cr:ZnS power amplifier also pumped by an EDFL [6,7]. The lasers operate at a repetition rate ($f_{rep}$) of 80 MHz, central wavelength of 2.35 μm, and FWHM bandwidth of 280 nm. Depending on the amplifier pump power, the average output power varies from 0.86 to 3.15 W with the pulse duration ranging respectively from 33 to 25 fs.

For the carrier envelope offset frequency ($f_{ceo}$) stabilization, a portion of the oscillator power is deflected with a beam splitter and focused into a periodically poled lithium niobate (PPLN) crystal (Fig. 1, inset). A custom-design PPLN with three sections of different QPM periods generates $2^{nd}$, $3^{rd}$ and $4^{th}$ harmonics that are used for $f_{ceo}$ detection via $3f$-to-$4f$ nonlinear interferometry at λ≈0.7 μm. The error signal is fed into a feedback loop that controls $f_{ceo}$ through changing the pump power of the Cr:ZnS oscillator. For the optical referencing of both combs, we utilize the second harmonic of the output, which is transmitted through a dichroic mirror of the oscillator cavity and heterodyned with a stable ultra-narrow linewidth 1064-nm laser. The beat signal $f_B$ is fed into a feedback loop to control the oscillator roundtrip length with a piezo transducer. The measured $f_{ceo}$ and $f_B$ offsets are phase-locked to synthesized radiofrequency (RF) signals referenced to a Rb clock. The integrated (10 Hz –10 MHz) phase noise of the $f_B$ and $f_{ceo}$ signals were <0.1 and <0.05 rad, respectively, which indicates robust phase locking. We consider the mutual coherence time between the two combs to be at least 100 s [28], and the absolute position of each comb tooth given by the Rb clock accuracy ($10^{-10}$).

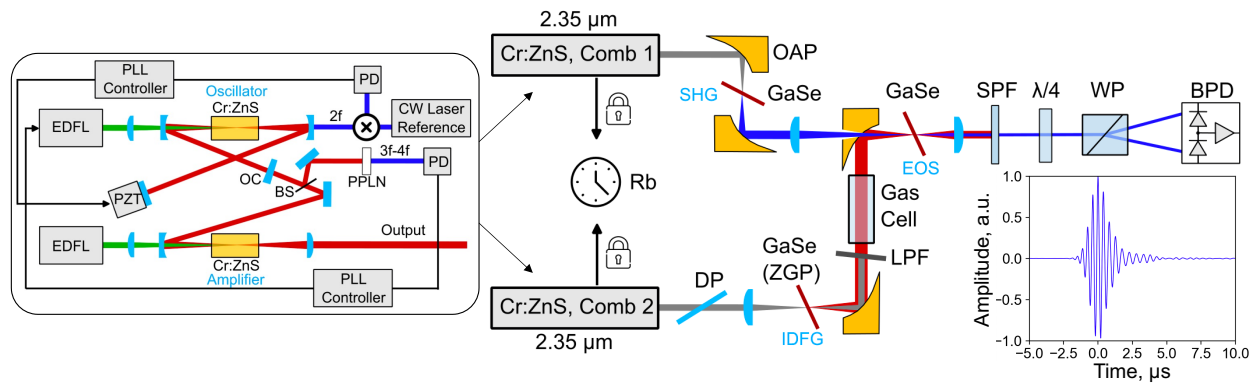

Fig. 1 Schematic of the EOS DCS spectroscopy setup. Inset (left) shows the Cr:ZnS laser system and its stabilization setup. Inset (right) shows the central portion of the interferogram with the temporal axis in the laboratory frame. EDFL, Er-doped fiber laser; PLL, phase-locked loop; PZT, piezo transducer; OC, output coupler; BS, beam splitter; PD, InGaAs detector for $f_B$ detection, and Si avalanche photodetector for $3f$-to-$4f$ interferometry; Rb, Rubidium atomic clock; DP, dispersive plate; OAP, off-axis parabolic mirror; LPF, longpass filter; SPF, shortpass filter; λ/4, quarter-wave plate; WP, Wollaston prism; BPD, InGaAs balanced photodetector.

### B. Mid-IR to THz combs produced by intrapulse difference frequency generation.

Generation of broadband transients via intra-pulse difference-frequency generation (IDFG) with few-optical-cycle pulses is a relatively simple but powerful technique for generating offset-free combs in the mid-IR – THz regions [31]. Depending on the application, we performed IDFG using two nonlinear crystals: ZnGeP$_2$ (ZGP) and GaSe. The ZGP crystal allows generation of LWIR transients with conversion efficiency exceeding 10% (thanks to its high nonlinearity and an excellent group-velocity matching between the 2.35-μm pump and IDFG output [16]). However, the spectral span of the output is limited to λ<12.5 μm, given by the ZGP transmission cut-off. In contrast, GaSe crystal can produce outputs spanning well beyond 25 μm, but with lower output power.

In the case of ZGP, the driving laser beam is pre-chirped with a 1-mm-thick sapphire plate (having the opposite sign of the group velocity dispersion) and focused into a 3 mm thick antireflection (AR) coated ZGP crystal using an $f = 75$ mm CaF$_2$ lens. The ZGP crystal is cut for type I phase matching with θ=51° and φ=0°. Since IDFG process requires pump with two orthogonal polarizations ('*o*' and '*e*' waves), dictated by the phase matching, the crystal is first rotated by 45° around the beam direction so that the original (horizontal) laser polarization will have both '*o*' and '*e*' components. The final crystal orientation was fine-tuned to produce the highest output power in the spectral region of interest and the generated LWIR beam was collimated with an off-axis parabolic mirror (OAP). With 2.8 W of the driving laser power the IDFG output power reached 300 mW after a 6.7 μm longpass filter (LPF). The integrated (10 Hz-10 MHz) LWIR intensity noise was measured to be 0.18%.

For IDFG in GaSe crystal, the driving laser beam was also pre-chirped with 1-mm sapphire plate and focused into a 1.3 mm thick GaSe crystal using an $f = 75$ mm CaF$_2$ lens. The crystal orientation is given by the type I phase matching with θ=11.3° and φ=90° (inset to Fig. 2a). Using 3.15 W of the driving laser power allows generation of 5 mW of the LWIR power after a 7.4 μm LPF. Fig. 2a represents the phase-matching function for IDFG, which shows that THz waves can be generated concurrently with LWIR waves at the same crystal orientation. Strictly speaking, thinner GaSe crystals are required to create broadband THz output, but this is outside the scope of this paper.

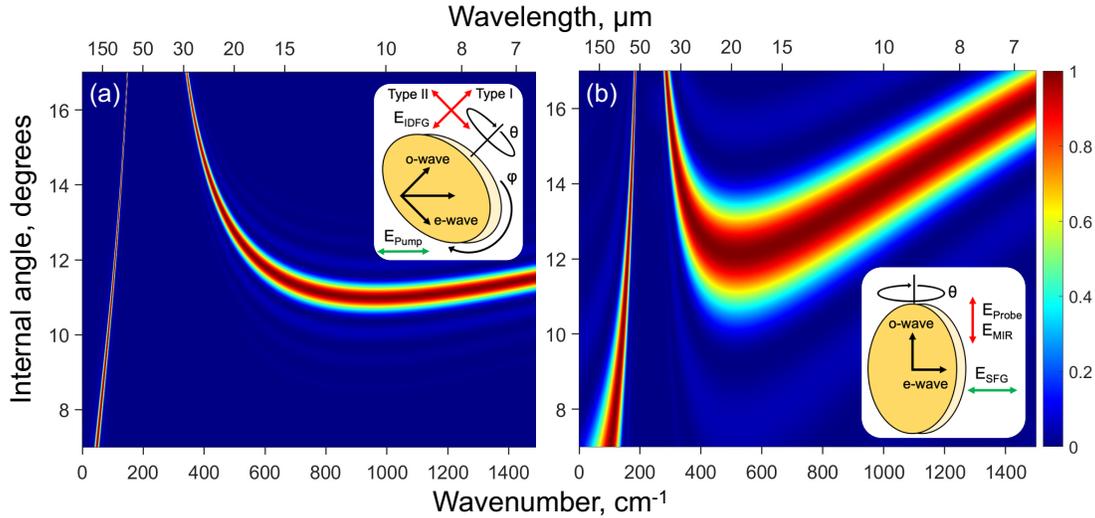

Fig. 2. (a) Phase-matching function |sinc(ΔkL/2)| of a 1.3-mm-thick GaSe crystal with respect to the generated IDFG output (x-scale) for the driving laser centered at 2.35-μm wavelength and the internal GaSe phase-matching angle θ (y-scale). The inset shows the GaSe crystal orientation. (b) Electro-optic phase-matching function |sinc(ΔkL/2)| of a 150-μm-thick GaSe with respect to the sampled IDFG wavelength (x-scale) for the EOS probe pulse centered at 1.15 μm. The inset shows the GaSe crystal orientation. The arrows indicate polarizations of interacting waves.

**C. Electro-optic sampling in the dual-comb configuration**

In the electro-optic sampling (EOS) method, the electric field of the mid-IR/THz transient is detected via induced change of the polarization of the near-infrared probe pulse inside an electro-optic (EO) crystal, where the two beams overlap. This polarization change is detected via ellipsometry using a balanced NIR photodetector. The key advantage of the DCS-EOS modality is an ultrabroadband (mid-IR to THz) spectral coverage with a single NIR detector, without the need for cryogenically cooled photodetectors [29,31].

The second 2.35-μm comb in our setup (Fig. 1) is used to generate a few-cycle NIR probe pulse. The laser output is focused into a 32-μm-thick GaSe crystal using an OAP mirror to produce the second harmonic (SH). A special care is taken to filter out the SH parasitically produced inside the laser cavity via random phase matching, to avoid distortion of the EOS signal. The SH GaSe crystal is oriented for the type I phase matching with θ=19.8° and φ=90° orientation; its thickness is chosen to utilize the full bandwidth of the driving laser in the SH generation process and to avoid the effect of the spatial walk-off. The generated probe has a central wavelength of 1.15 µm and 39 THz bandwidth at -10dB level. We did not measure the SH pulse duration but based on the width of the spectrum, it is expected to be close to 20 fs. With 2 W of the pump power the SH output power was 70 mW with vertically polarized beam and high (>99%) degree of polarization.

Next, both the probe and LWIR beams are spatially overlapped using an OAP mirror with a through hole (Fig. 1) and focused into yet another (EOS) GaSe crystal with the thickness of 150 µm orientated at $\theta=13°$ and $\varphi=90°$ for type I phase-matching (inset to Fig. 2b), corresponding to the sum (LWIR+NIR) frequency generation (SFG). Since the LWIR beam has a 45° polarization with respect to the probe beam, only its vertical projection (*o*-wave) participates in the nonlinear interaction while the horizontal component does not affect SFG and thus the balanced detector signal. Fig. 2b shows the phase-matching function for the 150-µm-thick EOS GaSe.

After GaSe crystal (Fig.1), the probe pulse goes through a short-pass filter (SPF, $\lambda<1100$ nm) to improve the dynamic range and the signal-to-noise ratio (SNR) of EOS by increasing the share of the SFG signal that carries the spectral information with respect to the total power of the NIR probe. Next, the beam is sent to an ellipsometry setup consisting out of quarter-wave plate, Wollaston prism and an InGaAs balanced photodetector (Thorlabs PDB450C, bandwidth 45 MHz). An attenuation wheel is used to keep the total power in each detector just below saturation (~1 mW). The differential signal (interferogram) from the balanced detector is radiofrequency filtered, digitized with a 16-bit analog-to-digital converter, coherently averaged, Fourier transformed, and frequency up-scaled to obtain the LWIR spectrum.

Shown in Fig. 3a is the spectrum obtained with GaSe as an IDFG crystal. One can see that the GaSe crystal produces a broad comb spanning 650 cm$^{-1}$ (9-22 µm) at -40db level. The spectral feature at 667 cm$^{-1}$ corresponds to the carbon dioxide absorption in the air. Interestingly, one can see a dip at ~510 cm$^{-1}$ (15.3 THz) related to the two-phonon lattice absorption of GaSe. Similarly, Fig. 3b depicts the spectrum obtained with ZGP as an IDFG crystal. Here we used two different phase-matching angles for the EOS GaSe crystal. The two different spectral contours indicate that the EOS detection bandwidth was not high enough to capture the whole IDFG spectrum. The absorption peaks at >1300 cm$^{-1}$ are due to water absorption in the surrounding air and sharp peaks near 900-1000 cm$^{-1}$ are due to isoprene absorption in the optical gas cell.

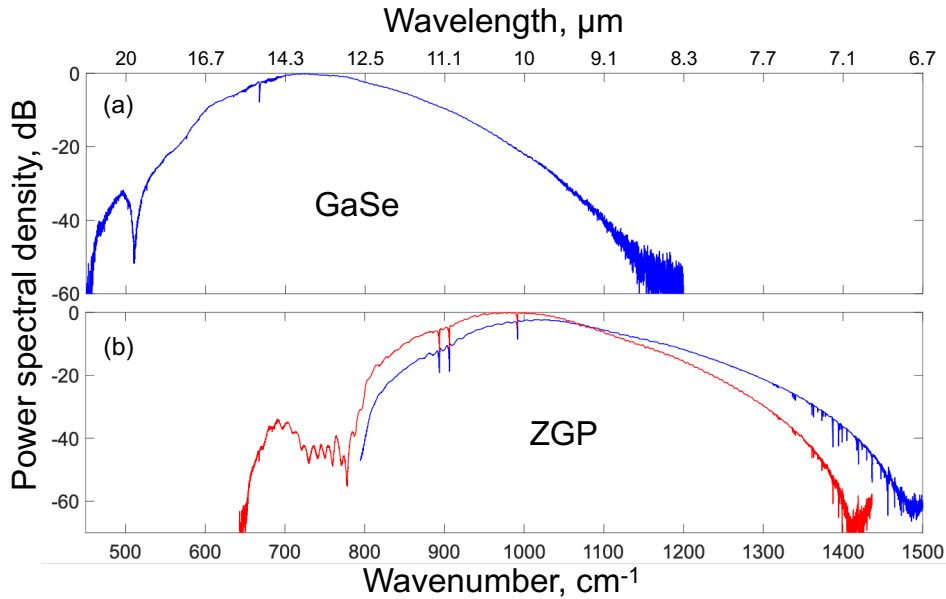

Fig. 3. (a) The spectrum obtained with GaSe as an IDFG crystal. (b) Spectra obtained with ZGP as an IDFG crystal for two different phase-matching angles of the EOS GaSe.

**III. HIGH RESOLUTION FIELD-RESOLVED DUAL COMB SPECTROSCOPY OF MOLECULES**

In this section we present our high-resolution spectroscopy study of several molecules that play an important role in exobiology and medical breath analysis. According to the DCS method, molecular vibrations are excited by few-cycle LWIR pulses and subsequently emit (in the same direction) coherent electric field detected through EOS. The nominal spectral sampling step is determined in our setup by the comb-mode spacing (80 MHz). However, when the absorption linewidths (predominantly Doppler-broadened in our case) are less than this spacing, we use the method of spectral interleaving, i.e., we combine the spectra taken with progressively shifted combs, in which case the spectral resolution can be well below 80 MHz [33]. The absorbance spectra for molecules (defined as $A=-\ln(I/I_0)$, where $I$ is the spectral intensity of a gas-filled cell, and $I_0$ is that of an empty cell) were obtained by normalizing the 'sample' spectrum to the one taken with vacuum in the cell.

### A. Mixture of $CO_2$ and $C_2H_2$

We started with taking high-resolution EOS-DCS absorption spectrum of a low-pressure mixture of carbon dioxide ($CO_2$) and acetylene ($C_2H_2$), as shown in Fig. 4. The two combs operated at $f_{rep}$=80 MHz with the repetition frequency offset between the combs of $\Delta f_{rep}$=91 Hz. A 20 cm-long absorption cell with antireflection (AR) coated (7-12 μm) Ge windows was filled with the gas mixture with concentration for each molecule of around 1% in $N_2$ buffer gas at 4 mbar total pressure. As a sensing comb, we used the IDFG comb produced in GaSe with the spectrum similar to the one shown in Fig. 3a. The absorption spectrum shown in Fig. 4a is combined from 8 interleaved comb-mode-resolved spectra, which allows us to fully resolve the narrow (66 MHz) absorption features. The expanded views of separately $CO_2$ and $C_2H_2$ spectra and their comparison with the HITRAN simulation (shown as inverted peaks) are depicted in Figs. 4b and 4c, respectively. Figure 4d illustrates the combination of eight distinct spectra obtained using shifted combs, each represented by points of varying colors. These spectra were combined to create a unified high-resolution spectrum, with the average spacing of 10 MHz.

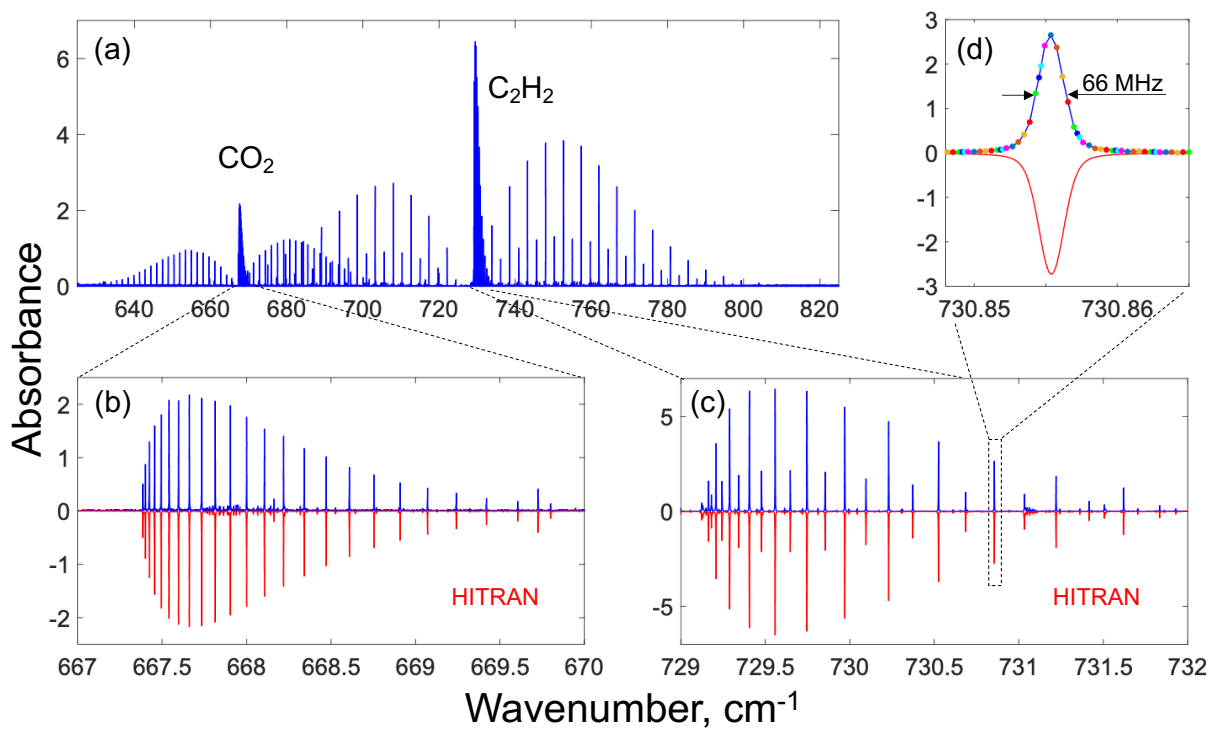

Fig. 4. (a) High-resolution spectrum of a mixture of $CO_2$ and $C_2H_2$ molecules with $N_2$ as a buffer gas at 4 mbar pressure. (b) Expanded view of the $CO_2$ spectrum and its comparison with the HITRAN simulation. (c) Expanded view of the $C_2H_2$ (acetylene) spectrum and its comparison with the HITRAN simulation. (d) Zoomed-in absorption line showing how spectral data points corresponding to shifted combs were combined in one spectrum.

### B. Methanol

High-resolution spectrum of methanol with predominantly Doppler-broadened linewidth (76 MHz) is shown in Fig. 5. In this and subsequent experiments, the two combs were operated with the repetition frequency offset of $\Delta f_{rep}$=69 Hz. A 45 cm-long absorption cell with AR coated Ge windows was filled with methanol vapor at 0.81 mbar pressure. As the sensing comb, we used the IDFG output from ZGP crystal. The absorption spectrum of Fig. 5 is a combination of 11 interleaved comb-mode resolved spectra. The simulated (HITRAN) spectrum is shown in red and inverted for clarity.

It can be seen from Figs. 4-5 that there is an excellent agreement with the HITRAN simulation for $CO_2$, $C_2H_2$, and methanol in terms of line positions and line widths.

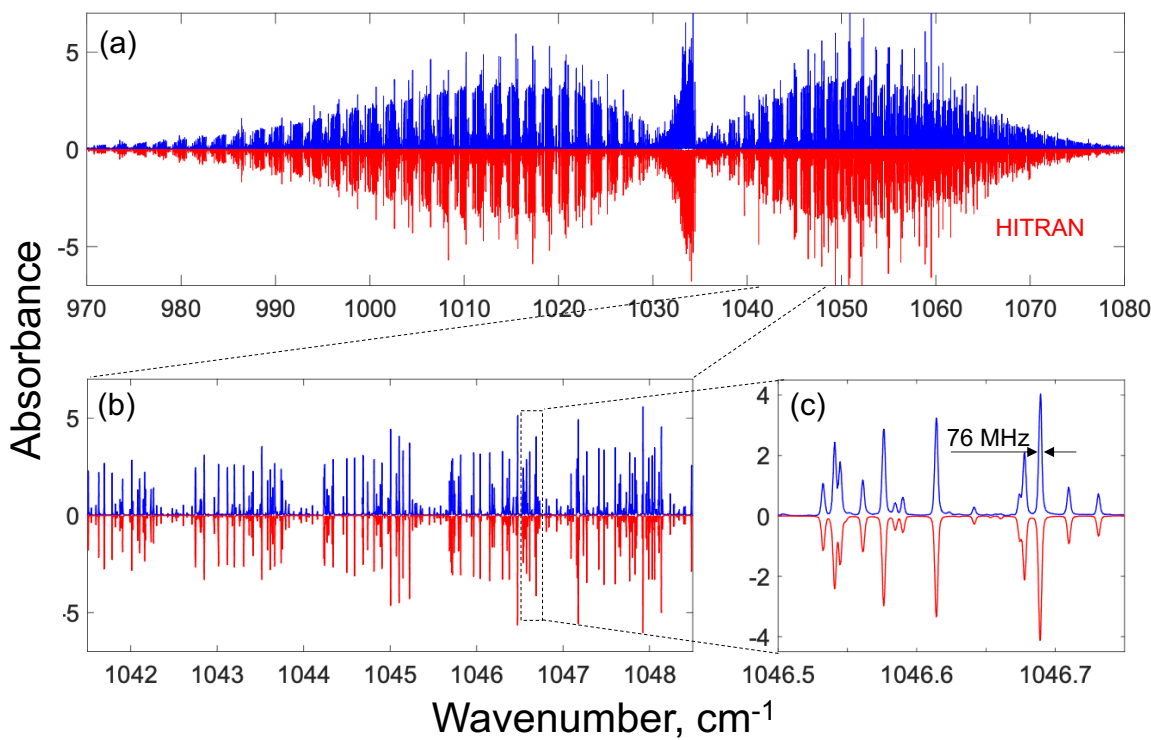

Fig. 5. (a) High-resolution LWIR spectrum of methanol at 0.81 mbar pressure. (b)-(c) Zoomed-in portions of the spectrum. The simulated (HITRAN) spectrum is shown in red and inverted for clarity.

**C. Ethanol, isoprene, and dimethyl sulfide (DMS)**

Figs. 6-8 show high-resolution vapor phase spectra ethanol, isoprene, and dimethyl sulfide obtained with the sensing comb produced via IDFG in ZGP crystal. We note that high-resolution spectra (better than 0.1 cm$^{-1}$) for these three molecules were not available in the literature or databases at the time of conducting our experiment. We employed a 45-cm-long absorption gas cell for this experiment; the vapor pressures of ethanol, isoprene, and dimethyl sulfide were set at 2, 4, and 13 mbar, respectively. We obtained 6 interleaved spectra for ethanol, 3 for isoprene, and 3 for dimethyl sulfide, such that the data point spacing varied between 13 and 26 MHz (0.0004–0.0009 cm$^{-1}$). We emphasize that the fine spectroscopic features for these three molecules had remained unexplored until now. A separate work will be dedicated to conducting a comprehensive spectroscopic study of these molecules, involving in-depth analysis and the creation of line lists for databases.

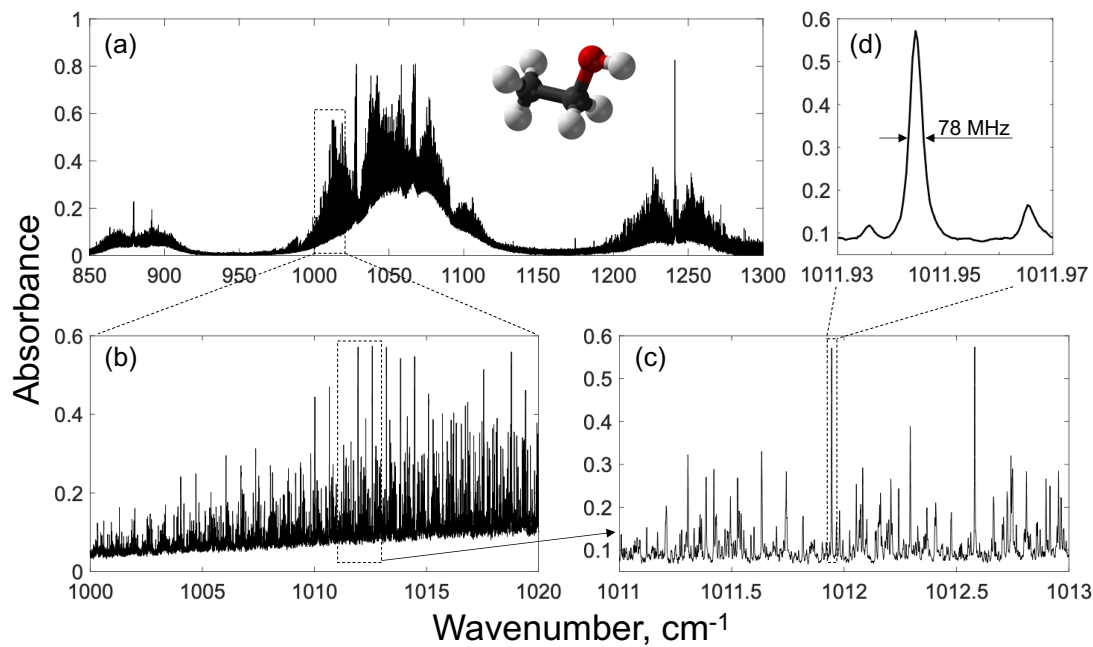

Fig. 6. High-resolution LWIR spectrum of ethanol at 2 mbar pressure. (b) –(d) Zoomed-in portions of spectrum.

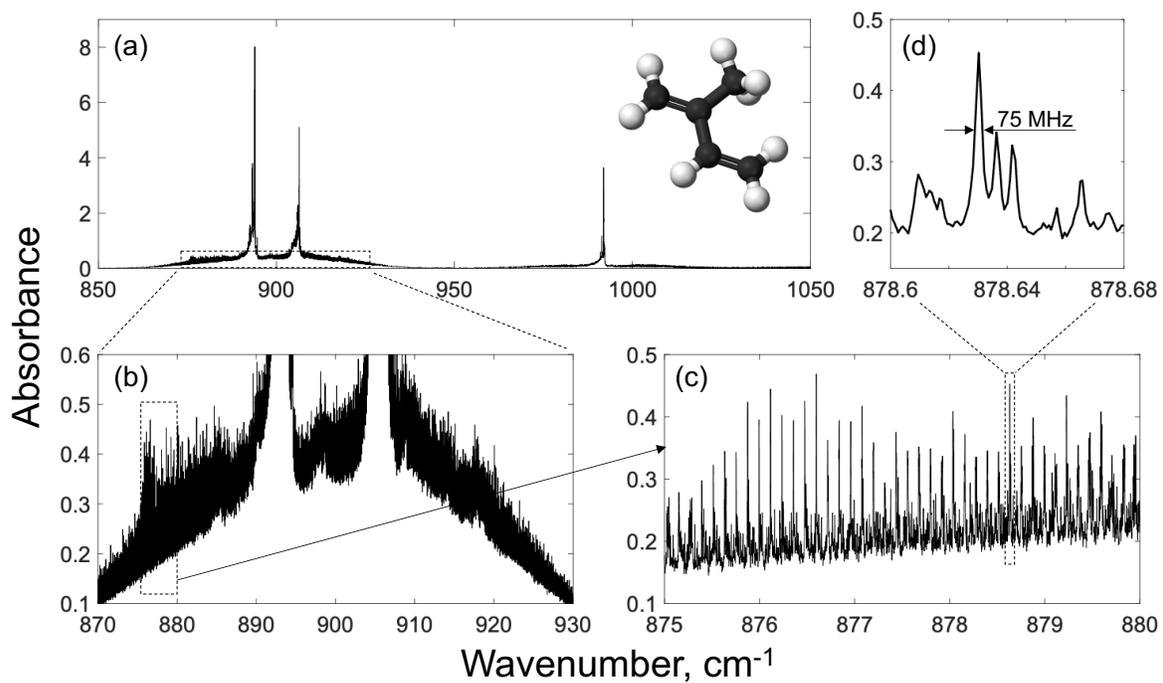

Fig. 7. High-resolution LWIR spectrum of isoprene at 4 mbar pressure. (b) – (d) Zoomed-in portions of spectrum.

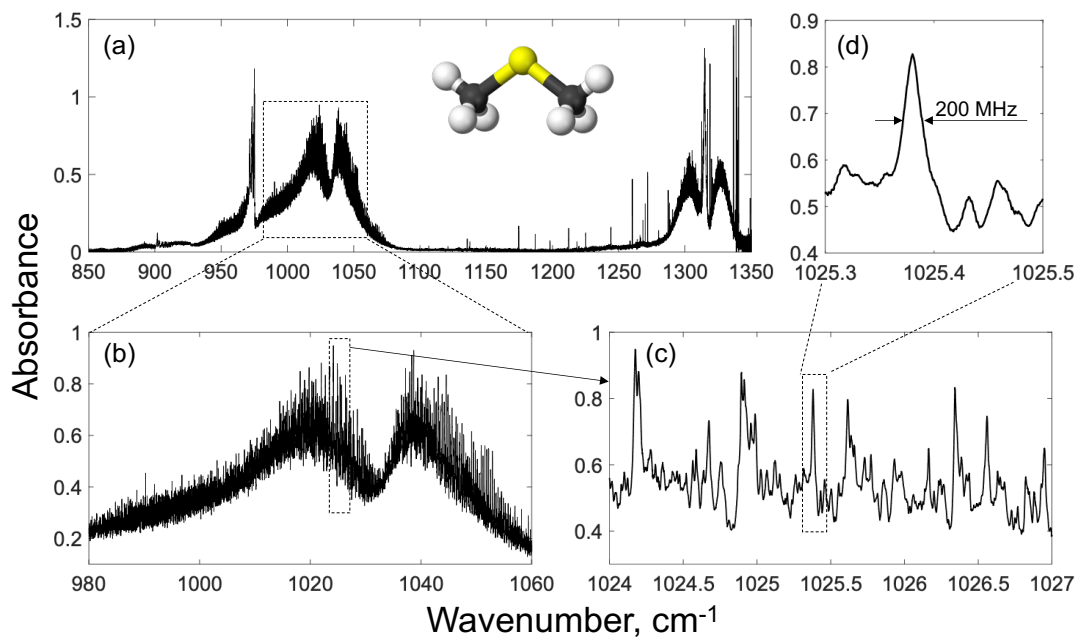

Fig. 8. High-resolution LWIR spectrum of DMS at 13 mbar pressure. (b) – (d) Zoomed-in portions of spectrum.

## IV. MEASUREMENTS AT VIDEO RATE

Thanks to the high SNR of our DCS-EOS system, we explored the capability of performing high-speed broadband LWIR spectroscopy. As the sensing comb we used the emission spectrum produced via IDFG in ZGP consisting of approximately 200,000 comb modes, with the comb center frequency around 1000 cm$^{-1}$ and comb width of 530 cm$^{-1}$ at -20 dB level. A 45-cm-long single-pass optical gas cell was filled with methanol vapor at a partial pressure ~ 1 mbar diluted in air at a total pressure of 20 mbar. Fig. 9 shows a portion of the spectrum of methanol at different acquisition times. Even at 0.0145-s acquisition time (69 Hz rate, single interferogram), we were able to detect the fine structure of methanol with the signal-to-noise ratio of 22. We did not use time-domain signal apodization, hence the full 80 MHz (comb-mode resolved) spectral resolution was preserved here.

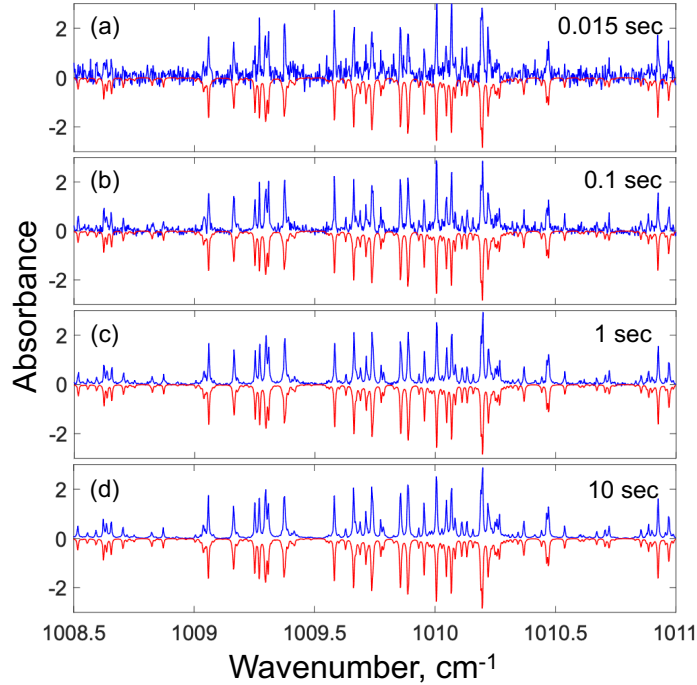

Fig. 9. Portion of the absorption spectrum of methanol between 1008.5 and 1011 cm$^{-1}$ corresponding to different averaging times ranging from 0.0145 s (single interferogram) to 10 s (687 interferograms). A 45-cm-long optical gas cell was filled with methanol vapor at a partial pressure ~ 1 mbar diluted in air at a total pressure of 20 mbar.

## V.  SPECTROSCOPY AT TERAHERTS (2-15 THz) FREQUENCIES

To demonstrate our system's ability to measure spectra at wavelengths >20 µm, we first adjusted the phase-matching angle of the EOS GaSe to detect a longwave portion (320 – 500 cm$^{-1}$, 9.6–15 THz) of the IDFG spectrum produced in GaSe (Fig. 10a). Since this spectral range contains numerous prominent water absorption lines in the surrounding atmosphere, we did not use the optical gas cell. Figure 10c displays the absorbance spectrum in this region (obtained by taking the negative natural logarithm of the emission spectrum and subtracting the baseline). The spectrum was compared with the HITRAN simulation (displayed in red color and inverted), revealing a good agreement between the two.

Similarly, we tuned the EOS GaSe crystal (by reducing its phase-matching angle θ) for the field detection in the terahertz region, below the GaSe crystal's strongly absorbing Reststrahlen band at about 5–10 THz (Fig. 10b). Despite of the fact that the thicknesses of both IDFG and EOS GaSe crystals were not optimized for terahertz generation and detection, we were able to observe a noticeable band at 1.5-5 THz. In order to verify that the observed band was not an artifact, we derived the absorbance spectrum from this band, as depicted in Figure 10d, and then compared this spectrum with the HITRAN simulation (displayed in red color and inverted) for absorption in ambient air predominantly contributed by water. We observed an agreement between the observed and simulated THz absorption peaks, with the exception that some of our measured peaks were saturated due to a lower signal-to-noise ratio in this particular region of the spectrum.

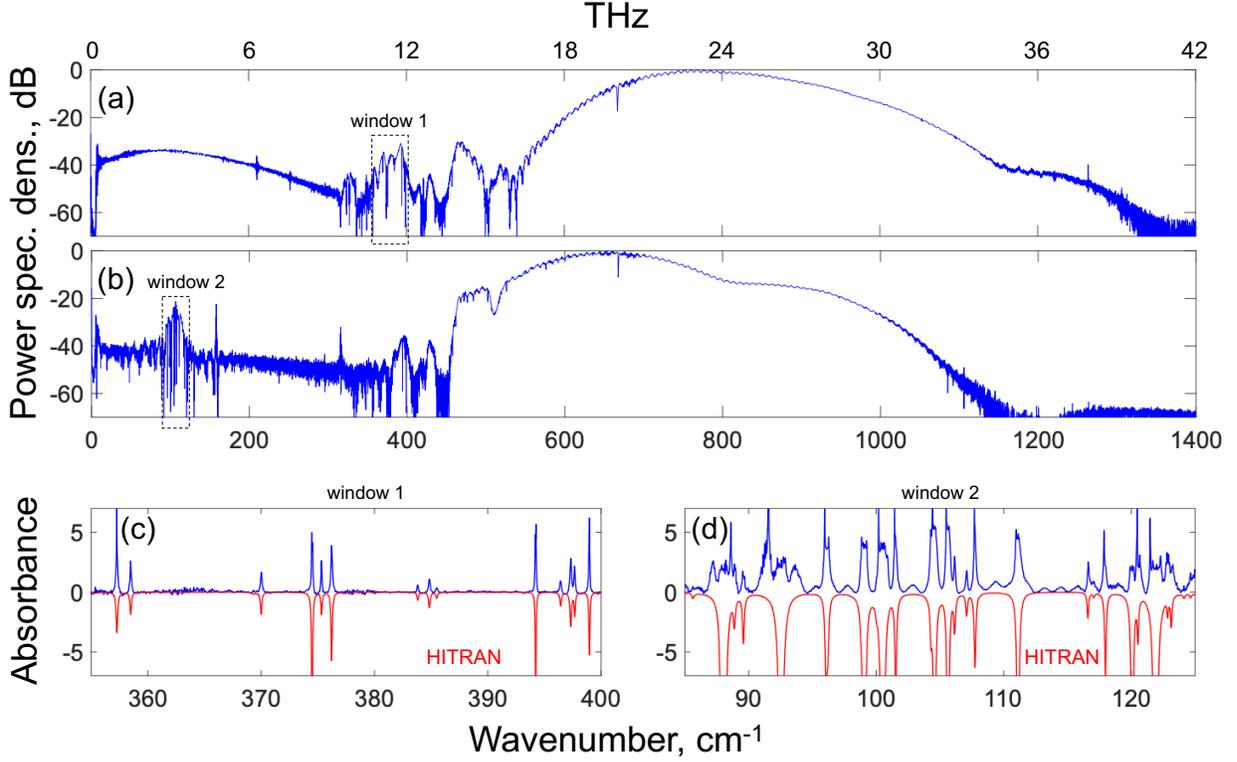

Fig. 10. (a) EOS-DCS spectrum that features atmospheric absorption in the 300-500 cm$^{-1}$ (9-15 THz) region. (b) EOS-DCS spectrum featuring atmospheric absorption lines below 5 THz. (c) Absorbance spectrum derived from the window 1 of the emission spectrum shown in (a). The simulated (HITRAN) spectrum for the ambient air is shown in red and inverted for clarity. (d) Absorbance spectrum derived from the window 2 of the emission spectrum shown in (b). The simulated (HITRAN) spectrum for the ambient air is shown in red and inverted.

## VI. DETECTION LIMITS AND THE FIGURE OF MERIT

In the time domain, our EOS-DCS signal exhibits a signal-to-noise ratio (SNR) of typically 1.2x10$^3$ for a single interferogram (we define SNR as a ratio of the peak signal to its standard deviation at the end of the interferogram, where we assume the field from the molecular free induction decay is negligible). The SNR scales as the square root of the number of averages, as verified in our experiments with up to 10$^6$ averages.

Based on our estimation this SNR is determined by the shot noise of the NIR balanced photodetector. In fact, when operating near the nominal saturation power of 1 mW for each detector channel (corresponding to $I_0 \sim 0.8$ mA detector current), we find that for the 45-MHz detector bandwidth and with the peak differential current $\Delta I \approx 0.25 \times I_0$ observed in our experiment, the shot noise-limited SNR for a single interferogram is approximately 1.3x10$^3$ that closely matches our experimental SNR. Considering the fact that SNR scales as the square root of the averaging time, for the comb span of ~ 500 cm$^{-1}$ we can achieve the $E$-field dynamic range of 10$^6$, corresponding to the intensity dynamic range of 10$^{12}$, in ~150 min of averaging.

In the frequency domain, by measuring the spectral power noise $\sigma_s$ vs. the averaging time $\tau$ in our experiment (we define $\sigma_s$ as the fractional standard deviation of spectral power density near the spectral maximum), we find that the spectral SNR (=1/$\sigma_s$) scales as $\sqrt{\tau}$, such that SNR/$\sqrt{\tau}$=62 Hz$^{1/2}$. For the number of modes M ≈ 200,000 within the central (-20 dB level) 530 cm$^{-1}$-wide portion of our typical LWIR comb (Fig. 3), this gives the DCS figure of merit (defined in [2]) of M×SNR/$\sqrt{\tau}$ = 1.2×10$^7$ Hz$^{1/2}$. This figure of merit surpasses the best reported value in the LWIR range of 7.3×10$^6$ Hz$^{1/2}$ [28] and provides a strong argument in favor of the EOS-DCS modality, especially when operated at longer wavelengths.

## VII. CONCLUSION

In summary, using EOS-DCS modality we performed high-resolution spectroscopic measurements across an ultra-broadband, 1.5 to 45 THz (6.6–200 µm), longwave frequency range with acquisition of octave-wide spectra with 200,000 comb-mode resolved lines at video rate (69 Hz). This result was facilitated by utilizing, as the driving source, mode-locked 2.35-µm Cr:ZnS lasers with the benefits of low noise and the ability to provide high (up to more than 10%) IDFG power conversion efficiency. With the nominal spectral resolution given by the comb-mode spacing (80 MHz, 0.0027 cm$^{-1}$), we were able to perform measurements with better than 10 MHz resolution via spectral interleaving. Also, we demonstrated our system's ability to do simultaneous measurements in the LWIR and THz domains. This opens up numerous possibilities for applications in fundamental spectroscopy, such as simultaneously studying absorption strengths of mid-IR and THz bands within the same experiment, allowing for cross-linking of molecular information. It also paves the way for creating highly accurate molecular spectroscopic databases and enables real-time medical diagnostics through multi-species exhaled breath analysis. Our next step will be compressing the driving 2.35-µm pulses to sub-10 fs duration that will allow extending the spectral coverage to the whole 1–100 THz range.

**Acknowledgements**

We acknowledge support from the Defense Advanced Research Projects Agency (DARPA), grant number W31P4Q-15-1-0008; from the Office of Naval Research (ONR), grants numbers N00014-15-1-2659, N00014-18-1-2176, N00014-17-1-2705, and N68335-20-C-0251; and from the Department of Energy (DOE), grant number B&R #KA2601020; US Air Force Office of Scientific Research (AFOSR), grant number FA9550-23-1-0126.